\documentclass[%
 aip,
 amsmath,amssymb,
 reprint,%
]{revtex4-2}

\usepackage{graphicx}
\usepackage{dcolumn}
\usepackage{bm}

\usepackage[utf8]{inputenc}
\usepackage[T1]{fontenc}
\usepackage{mathptmx}
\usepackage{etoolbox}

\makeatletter
\def\@email#1#2{%
 \endgroup
 \patchcmd{\titleblock@produce}
  {\frontmatter@RRAPformat}
  {\frontmatter@RRAPformat{\produce@RRAP{*#1\href{mailto:#2}{#2}}}\frontmatter@RRAPformat}
  {}{}
}%
\makeatother
\begin{document}

\preprint{AIP/123-QED}

\title{Chemical interactions in active droplets}
\author{Prateek Dwivedi}
\author{Sobiya Ashraf}%
 

\author{Pawan Kumar}

\author{Dipin Pillai}

\author{Rahul Mangal$^*$}
\email{mangalr@iitk.ac.in}
\affiliation{%
Department of Chemical Engineering, Indian Institute of Technology Kanpur, India
}

\date{\today}

\begin{abstract}
Interactions among biologically active agents is facilitated by their self-generated chemical and hydrodynamic fields. In order to elucidate the pair-wise interactions between such micro-organisms, we employ active droplets as a model system, capable of self-generating chemical and hydrodynamic fields. We demonstrate that the solute Péclet number ($Pe$), characterizing the relative strength of its convective to diffusive transport, plays a crucial role in determining how the chemical and hydrodynamic fields impact their interactions. Our findings reveal that at low $Pe$, the interaction is predominantly governed by chemo-repulsive interactions, leading to droplets avoiding physical contact. Conversely, at elevated $Pe$, it is the hydrodynamic interactions that dictate the dynamics, leading to physical engagement. However, irrespective of $Pe$, the interaction of a droplet with the chemical trail of another droplet is always governed by chemo-repulsive effects. Furthermore, our results establish that the chemo-repulsive deflection/rebounding of droplets is primarily decided by the droplets' inherent chemical polarity, as determined by its $Pe$, independent of their approach orientation. Our findings provide a robust methodology for optimizing the outcomes of binary interactions among chemically active droplets, laying the groundwork for future investigations into their collective dynamics.
\end{abstract}

\maketitle


\section{\label{sec:level1}INTRODUCTION}

Artificial micro-swimmers use anisotropic interactions with their surroundings to achieve out-of-equilibrium self-propulsion. One of the popular routes is to use Janus particles with an intrinsic asymmetry in their physical/chemical morphology that facilitates a physico-chemical gradient in their surrounding, propelling them through a mechanism known as `phoresis' \cite{anderson1989colloid}. Depending on the exact mechanism behind the generation of the gradient, JPs are known to self-propel via self-diffusiophoresis, self-thermophoresis or self-electrophoresis. In the case of oil (water) droplets in water (oil), due to the lack of any inherent shape/composition asymmetry, an alternate chemical pathway is used to spontaneously generate asymmetry in interfacial tension ($\gamma$) gradient along the droplet interface. Among the various mechanisms that generate interfacial tension gradients, micellar solubilization is widely studied \cite{peddireddy2012solubilization}. In this mechanism, when a droplet is dispersed in a continuous phase with the surfactant concentration being much greater than the critical micellar concentration (CMC), the droplet solubilizes. During this process, free micelles from the bulk enter a thin solubilization zone surrounding the droplet, where they incorporate solubilized droplet phase molecules and by collecting additional surfactants, depart as oil-filled micelles. This solubilization process results in local fluctuations in surfactant density around the droplet, creating a spontaneous asymmetry in the interfacial tension gradient. This asymmetry triggers the onset of Marangoni stress at the interface, driving fluid flow from regions of low interfacial tension to regions of high interfacial tension along the droplet surface, therby causing the droplet to propel in the opposite direction. As the droplet propels, it encounters more empty micelles at its front and leaves a trail of filled micelles in its wake, thereby reinforcing the initial asymmetry in surfactant concentration at the droplet interface. This advective transport of surfactant/empty micelles couples non-linearly with the droplet velocity and concentration fields around the droplet, leading to sustained self-propulsion \cite{morozov2019nonlinear, michelin2023self,dwivedi2022self}. 


Using an axisymmetric 2D model accounting for combined diffusiophoretic and Marangoni effects, Morozov and Michellin demonstrated that a dimensionless $Pe$ governs the onset of self-propulsion. Beyond a critical value of $Pe=4$, a symmetry-breaking instability was shown to result in a front-aft asymmetry, leading to self propulsion. With increase in $Pe$, stronger advection resulting in chaotic oscillations due to the emergence of higher spherical modes that hinder the front-aft asymmetry \cite{morozov2019nonlinear}. Recent experiments have verified this prediction for swimming oil droplets in aqueous media by increasing the surfactant concentration \cite{izzet2020tunable}, adding molecular solutes \cite{dwivedi2021solute, hokmabad2021emergence}, or changing the droplet size \cite{suga2018self,suda2021straight}. Using macro-molecular additives, Dwivedi $et. al$ demonstrated that $Pe$ can be reduced resulting in more persistent motion compared to aqueous additive-free media. \cite{dwivedi2023mode} It was also shown that the lower $Pe$ was characterized by a puller mode, which explained the more persistent nature of trajectories. In another recent study, authors further demonstrated experimentally that on rendering viscoelastic properties to the continuous media, the droplet deforms from its usual spherical shape. Using a simple analytical model, they showed that the shape deformation was due to an excess elastic normal stress at the droplet interface generated by the deformed polymer chains \cite{dwivedi2023deforming}.\\

These active droplets have been shown to exhibt intriguing phenomena commonly observed in microbial motion, such as chemotaxis \cite{jin2017chemotaxis}, rheotaxis \cite{dey2022oscillatory, dwivedi2021rheotaxis}, and motion under gravity \cite{castonguay2023gravitational}. Apart from the fascinating characteristics displayed by isolated swimming droplets, their collective behavior has also garnered significant interest. The study of collective dynamics in these droplets offers valuable insights into the collective behavior observed in biological microswimmers, including flocking or swarming \cite{verstraeten2008living}, predator-prey interactions \cite{tang2020predator}, the formation of bacterial colonies \cite{nadell2008sociobiology}, and inter-cellular communications. Furthermore, the collective behavior of these artificial swimmers enables them to accomplish tasks that would otherwise be unattainable by isolated swimmers. For instance, recent studies have reported formation of dynamic self-assemblies that can evolve into new structures depending on upon the physical confinement or nature of activity \cite{thutupalli2011swarming}. Spontaneous rotation of such dynamic clusters has also been reported \cite{hokmabad2022spontaneously}. A few recent studies explored chemotactic interaction between an oil droplet in an aqueous surfactant solution and the chemical trail of another droplet \cite{hokmabad2022chemotactic}, and by varying the chemical structure and concentration of oils and surfactants were shown to mimic predator-prey dynamics \cite{meredith2020predator}.\\

The emergence of collective dynamics in these systems is a result of the pairwise interactions between the swimmers\cite{blaschke2016phase}. In biological swimmers, the interactions are mostly governed by chemical signalling and hydrodynamics mediated by the intermediate fluid \cite{peng2021imaging}. The hydrodynamic signature of a swimmer depends on its size, speed, physical confinement and more importantly on its swimming mode. For example, pushers are characterized by pushing fluid from both their front and rear ends while drawing it in from their equatorial region, whereas pullers do the reverse \cite{underhill2008diffusion}. In active droplets, besides the short-range hydrodynamic interactions, the chemical field generated by the swimmers play a vital role in their interactions. For micellar solubilization based active droplets, the trail of filled micelles released by the swimming droplets is generally avoided by other droplets. The transport of free surfactants toward regions rich in filled micelles is hindered by electrostatic repulsion, and therefore such regions serve as fuel-depleted domains for other droplets which they avoid - a phenomenon known as negative chemotaxis \cite{jin2017chemotaxis}. While chemical interactions between droplets have been documented, most of these studies have focused on solubilizing, but originally non-swimming, droplets. The droplet solubilization process can either be reaction-limited or mass-transfer limited. In reaction-limited solubilization process, the number of filled micelles produced per unit area is independent of droplet size, resulting in a constant rate of oil dissolution \cite{hokmabad2021emergence,dwivedi2021solute}. On the other hand, under diffusion-limited solubilization process, the dissolution rate is limited by the diffusivity of micelles/surfactants and thus depends on droplet size \cite{moerman2017solute,izzet2020tunable}. In the diffusion-limited micellar solubilization process, Moreman \emph{et al.} examined interactions between non-swimming droplets, demonstrating repulsive forces mediated by solutes released during solubilization \cite{moerman2017solute}. Similarly, in the reaction-limited micellar solubilization process, Zarzar and co-workers showed that by varying the chemical structure of the oil phase, surfactant structure, and surfactant concentration, both attractive and repulsive interactions can occur between non-swimming solubilizing droplets \cite{wentworth2022chemically}. Additionally, they investigated non-reciprocal interactions between solubilizing and non-solubilizing droplets (both non-swimming), showing that solubilizing droplets accelerate toward non-solubilizing ones \cite{meredith2020predator}. For solubilizing self-propelling droplets, Hokmabad \emph{et al.} demonstrated pairwise interactions with droplets swimming at moderate $Pe$ resulting in pusher mode. These investigations indicated strong hydrodynamic interactions between the droplets \cite{hokmabad2022spontaneously}. Recently, Lippera \emph{et al}.\, theoretically predicted repulsive pair-collisions between droplets when interacting through chemical fields at low $Pe$ \cite{lippera2021alignment,lippera2020collisions}. Despite these recent advancements, there is a lack of detailed experimental studies that provide a comprehensive understanding of pairwise interactions and the role of $Pe$, especially at low $Pe$, in micellar solubilization-based droplet self-propelled droplet systems, which we aim to address in this study. To the best of our knowledge, this is the first experimental investigation demonstrating that droplet-droplet interactions can be controlled by simply tuning the system's $Pe$. High $Pe$ droplets are generated in usual aqueous media wherein droplets propel as pushers. Lower $Pe$ propulsion was achieved by adding macro-molecules to the surrounding medium, wherein, droplets propel as pullers. The study reveals strong repulsive interaction among the droplets with lower $Pe$, leading to strongly scattering collisions. On the other hand, for higher $Pe$, the presence of stronger hydrodynamic interactions can lead to a variety of two-body interactions.

 \section{\label{sec:level1}MATERIALS AND METHODS}

\subsection{\label{sec:level2}Chemicals}

Using DI water, an aqueous solution containing 6 wt.$\%$ Tetradecyltrimethylammonium bromide (TTAB), a cationic surfactant obtained from Loba Chemicals, was prepared. In some experiments, 1 wt.$\%$ Polyethylene oxide (PEO) with a molecular weight of 8000 kDa, sourced from Sigma Aldrich, was added to the aqueous TTAB solution as a macromolecular solute. Droplets ($\sim$ 50 $\mu$m) of 4-Cyano-4'-pentylbiphenyl (5CB), a thermotropic liquid crystal acting as the oil phase, also procured from Sigma Aldrich, were prepared using a micro-injector (Femtojet 4i, Eppendorf) by injecting 5CB into the TTAB solution (with or without polymer solute).  Filled micelles rich aqueous TTAB solution was prepared by pre-dissolving 2 $\mu$$l$ of 5CB oil in 100 $\mu$$l$ of 6 wt.\% TTAB solution for almost 24 hrs. The number density of the droplets was kept $\sim$ 20 droplets per 10 mm$^{3}$ ($10^{10}$$\mu$m$^{3}$) of TTAB solution to observe droplet pairwise interactions. For multibody interactions, number density of the droplets was kept $\sim$ 250 droplets per 10 mm$^{3}$ ($10^{10}$$\mu$m$^{3}$) of TTAB solution For fluorescence experiments, Nile Red, an oil-soluble fluorescent dye was obtained from Sigma Aldrich. For Particle image velocimetry (PIV) experiments, fluorescent polystyrene particles ($\sim$ 500 nm) were procured from Thermo Fisher Scientific.  

\subsection{\label{sec:level2}Brightfield experiments}

The resulting emulsion was then injected into a custom-built Hele–Shaw optical cell with a vertical gap of 100 $\mu$m. The cell was prepared using glass slides cleaned through ultrasonication in ethanol, followed by plasma treatment and nitrogen drying. Considering gap thickness to be marginally larger than the droplet size, this setup restricted droplet movement mostly to a 2D X-Y plane (Fig. \ref{Figure1}A). The optical cell was mounted on a thermal stage at constant temperature of 25$^o$C on an upright microscope Olympus BX53. Droplets' motion was captured with Olympus LC30 camera at 2-10 frames/s in brightfield mode.

\subsection{\label{sec:level2}Fluorescence experiments}

The chemical field around the droplets was optically visualized by incorporating Nile Red into the 5CB oil droplets. To analyze the fluid flow around the droplets, Particle Image Velocimetry (PIV) was employed. In this technique, fluorescent polystyrene particles ($\sim$ 500 nm) were dispersed in the surfactant solution, which are convected by the flow-field generated by the swimming droplets. The optical setup includes mounting the Hele-Shaw cell on an Olympus IX73 inverted microscope, equipped with an Olympus U-RFL-T fluorescence illuminator (Mercury Burner USH-1030L) and a laser (wavelength approximately 560 nm) for dye excitation. A FLIR ORX-10G-71S7C-C camera (3208 x 2200 pixels), attached to the microscope with 10x magnification, and controlled via a computer interface (SpinView software), captured videos with consistent exposure settings. The velocity vectors around the droplets were determined using the PIVlab interface in MATLAB.

 \section{\label{sec:level1}RESULTS AND DISCUSSION}

\subsection{\label{sec:level2}Micellar solubization and swimming mode}
\begin{figure*}
\centering
\includegraphics[scale=0.45]{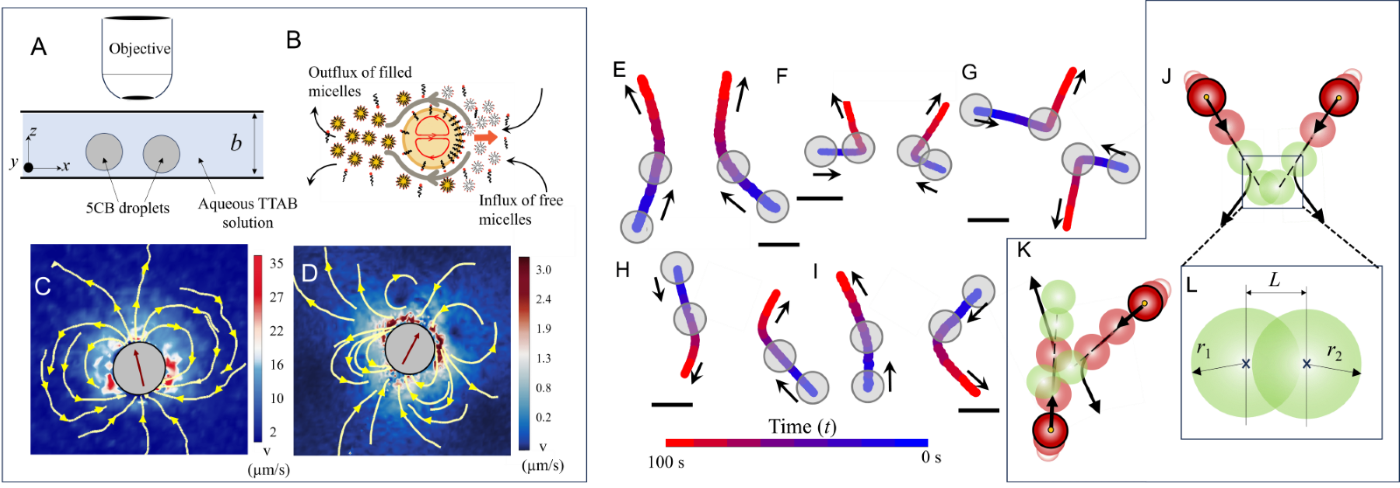}
\caption{ (A) Schematic illustrating the experimental setup. (B) Schematic illustrating the influx/consumption of surfactant micelles and the generation of filled micelles by an active droplet during self-propulsion. Flow-field generated by the self-propelling droplet (C) in aqueous TTAB solution where the droplet is a weak pusher and (D) in aqueous TTAB solution doped with 8000 $kDa$ PEO (c$_{PEO}$ = 1 wt$\%$) where the droplet is a puller. Color bar indicates the magnitude of velocity field around the droplet.(E-I) Representative trajectories of active droplets involved in binary interactions in 1 wt.$\%$ PEO (Polyethylene oxide 8000 $kDa$ molecular weight) aqueous 6 wt.$\%$ TTAB surfactant solution. Schematic showing the 2-D projected areas of approaching active droplets extrapolated with no deflections, for (J) the case of high overlapping (K) the case of no overlapping. (L) Schematic representing $r_{1}$ and $r_{2}$ as droplet radii and $L$ as distance between the droplet centers. Scale bars indicate a length of 50 $\mu$m.}
\label{Figure1}
\end{figure*}

We study the self-propelled motion of 4-n-pentyl-4\textquotesingle-cyanobiphenyl (5CB) droplets in aqueous Tetradecyltrimethylammonium bromide (TTAB) surfactant solution. The TTAB concentration is kept at 6 wt.\%, which is significantly above its critical micellar concentration (CMC$=0.13$ wt.\%). As a result, 5CB droplets undergo micellar solubilization-based self-propulsion in aqueous TTAB solution (cf. Fig. \ref{Figure1}B for a schematic of propulsion). The droplet solubilization rate ($k$) is 9.5 $n$m s$^{-1}$, with $l = \frac{D}{k}$ (where $D$ is the micelle diffusivity) estimated to be approximately 3000. Given that $\frac{l}{a} \gg 1$, where $a$ represents the droplet size, the solubilization process is determined to be reaction-limited \cite{moerman2017solute}. The droplets propel with random trajectories at an average speed, $V$ $\sim$ 35-40 $\mu$m s$^{-1}$. The swimming characteristics are strongly determined by the P\'{e}clet number $\left(Pe=\frac{aV}{D}\right)$ associated with their motion. Here, $a$ is the droplet size, $V$ is the experimentally measured time-averaged droplet speed, and $D$ is the diffusivity of micelles/surfactants. Following the protocol outlined by Hokmabad \emph{et al.} \cite{hokmabad2022chemotactic}, the diffusivity $D$ of filled micelles is calculated to be $\sim$ 25 $\mu$m$^{2}$s$^{-1}$, resulting in $Pe$ $\sim$ 75, and the droplet being a weak pusher.

Addition of Polyethylene oxide (PEO) as a macromolecular solute results in lower droplet speed, $\sim$ 1-3 $\mu$m s$^{-1}$, with more persistent trajectories. The rate of droplet solubilization remains unchanged. However, a disparity in length scales between the size of filled micelle and the characteristic polymer length scale results in the breakdown of continuum hypothesis-based prediction of diffusivity \cite{dwivedi2023mode}. As the characteristic polymer length scale greatly surpasses the size of diffusing micelles, micelle diffusivity, $D$, remains almost identical to that observed in PEO-free aqueous TTAB solution. Moreover, the increase in bulk viscosity (10 Pa.s) hinders fluid convection, thereby leading to a reduction in $Pe$ to $\sim$ 3. As a consequence of such low $Pe$, a puller swimming mode with more persistent motion is observed in the presence of a macromolecular solute (PEO). It is important to note here that no elastic effects are introduced in the ambient medium by the addition of additives. The fluid streamlines around the droplet for the case of a pusher and puller droplet are demonstrated using PIV micrographs in Fig. \ref{Figure1}C and D, respectively.

\subsection{\label{sec:level2}Interactions at low $Pe$}

After benchmarking the experimental setup, we first focus on the pairwise interactions between similarly sized ($\sim$50 µm) active droplets in a PEO+TTAB aqueous solution, corresponding to the low $Pe$ regime. To avoid multi-body (three or more) interactions, low number density of 5CB droplets is maintained. Through a series of careful experiments, we capture $\sim$ 40 pair interactions. In each of these cases, whenever a pair of droplets approach each other, they are found to scatter away after reaching a minimum distance ($d_\text{{min}}$). Fig. \ref{Figure1}E-K illustrate a few representative trajectories of such pair interactions, exhibiting some intriguing preliminary observations. In certain cases (see Fig. \ref{Figure1}E,F), a pair of droplets approach each other symmetrically before parting away along mirrored trajectories. 
In other instances, a different behavior is observed (Fig. \ref{Figure1}G), characterized by the droplets diverging in opposite directions in almost anti-mirrored trajectories. The key factor determining the fate of the trajectories is the time-lag in the approach of droplets. For instance, Fig. \ref{Figure1}F illustrates a scenario with little to no lag between the droplets as they approach each other, whereas, in Fig. \ref{Figure1}G, a more pronounced time-lag exists. With increasing time-lag, the droplets deflect with a further loss of symmetry between their trajectories, as depicted in Fig. \ref{Figure1}H,I. These findings highlight the critical role of time-lag in their interactions. To quantify time-lag, we determine the \emph{overlap percentage}, defined as $\psi = \frac{(r_{1}+r_{2})-L}{(r_{1}+r_{2})}\times100$. This metric estimates the maximum percentage of overlap between the 2D projected areas of the droplets, had they continued on their initial paths without any deflection. Here, $r_{1}$ and $r_{2}$ are the radii of the interacting droplets, and $L$ is the distance between their centers at the extrapolated point of maximum overlap, as shown in the schematic in Fig. \ref{Figure1}L. For droplets with no time-lag, the anticipated value of $\psi$ = 100 $\%$. As time-lag increases, $\psi$ decreases and reaches $0\%$, wherein droplets barely make contact when extrapolated. Cases with $\psi<0$, indicative of interactions wherein droplets do not overlap at all, are also assigned as $\psi=0\%$.

\begin{figure*}
\centering
\includegraphics[scale=0.4]{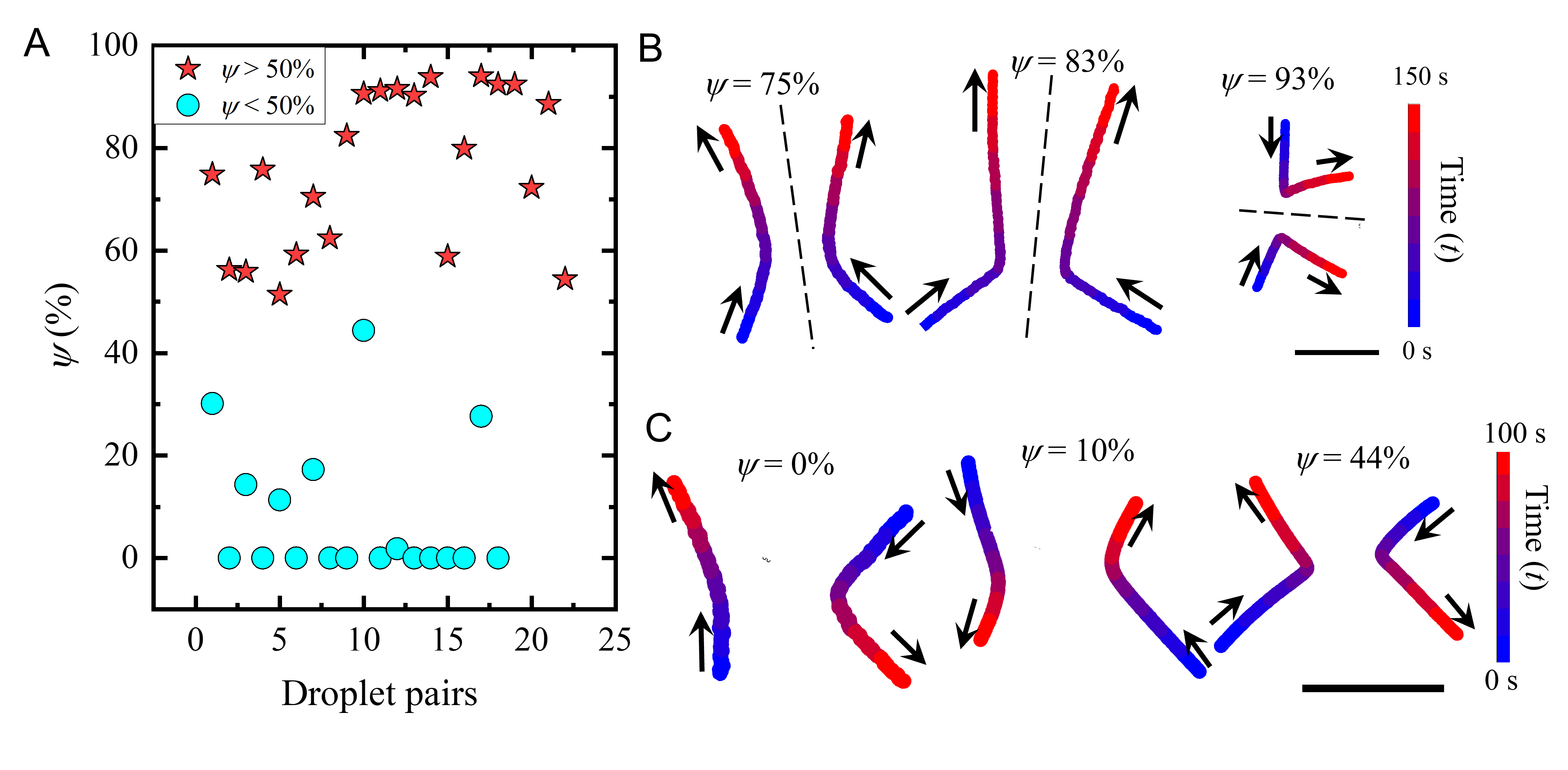}
\caption{(A) Overlap percentage ($\psi$) values for droplet pair interactions for the cases of $\psi>50\%$ and $\psi<50\%$. Representative trajectories of droplet pair interactions for the case of (B) $\psi>50\%$, dotted lines represent anticipated plane of reflection and (C)  $\psi<50\%$. Scale bars indicate a length of 100 $\mu$m.} 
\label{Figure2}
\end{figure*}

After assessing the $\psi$ values for all droplet interactions, we observe that instances with $\psi >$ 50 $\%$ (refer to Fig. \ref{Figure2}A) correspond to pair interactions where a droplet's deflection closely mirrors its partner's deflection. This reflection appears to occur about a plane passing through midpoint of their center-to-center line, as depicted in the trajectories shown in Fig. \ref{Figure2}B. Moving forward, we turn our attention to droplet interactions with $\psi<50\% $. In Fig. \ref{Figure2}A, the $\psi$ values for 18 such pair interactions are presented, and Fig. \ref{Figure2}C exhibits a few representative trajectories. Interestingly, several pairs show a $\psi$ value of 0\%, suggesting that, even without any potential overlap between the droplets, they still interact. It is crucial to note that $\psi=0\%$ also includes cases where one droplet interacts with the trail/wake of the other droplet, which we will discuss in later sections (refer to Fig. \ref{Figure5}E,F). Indeed, for all such droplet-wake interactions, $\psi$ fails to quantify the actual lag between the droplets. For now, our focus is limited to droplet-droplet interactions and all instances of droplet-wake interactions will be discussed later. 

Numerous previous studies have highlighted the significance of the interplay between flow field (hydrodynamics) and the concentration field (chemical field) of solutes, such as filled micelles, on the dynamics of chemically active droplets \cite{morozov2019nonlinear, hokmabad2021emergence, dwivedi2023mode, dwivedi2021rheotaxis}. Consequently, we anticipate that these factors play a crucial role in shaping the pairwise interaction of active droplets in our experiments. In the presence of PEO, since the droplets exhibit a puller swimming mode, they pull the bulk fluid in from their poles and push it out from their equatorial region. Given this hydrodynamic signature of puller droplets, we expect them to mutually attract, especially in head-on encounters. In fact, using the squirmer model, Ishikawa \emph{et al.} predicted a hydrodynamic attraction among puller swimmers, leading to a brief period of sustained contact before they separate again \cite{ishikawa2006hydrodynamic}. Further, several theoretical analyses suggest that hydrodynamic interactions among squirmers in puller mode lead to clustering \cite{zottl2014hydrodynamics,theers2016modeling,theers2018clustering}. Therefore, it must be noted that our findings of puller droplets repelling each other, regardless of their approach angle, contradict the behavior anticipated solely from pure hydrodynamic interactions.

In 2020, Lippera \textit{et al.} conducted a numerical investigation of head-on pair collisions of chemically active droplets \cite{lippera2020collisions,lippera2020bouncing}. The study predicted that irrespective of the underlying $Pe$, approaching droplets would consistently experience a chemical repulsion due to the accumulation of solute in their intermediate region. Subsequently, their numerical study specifically investigated the rebounding interactions of droplet pairs, focusing on the influence of chemical field  \cite{lippera2021alignment}. Our experimental observations at low $Pe$ (for all $\psi$) are in very good agreement with these numerical predictions, reinforcing the predominance of chemical field in the observed pair-interactions. 

\begin{figure*}
\centering
\includegraphics[scale=1.4]{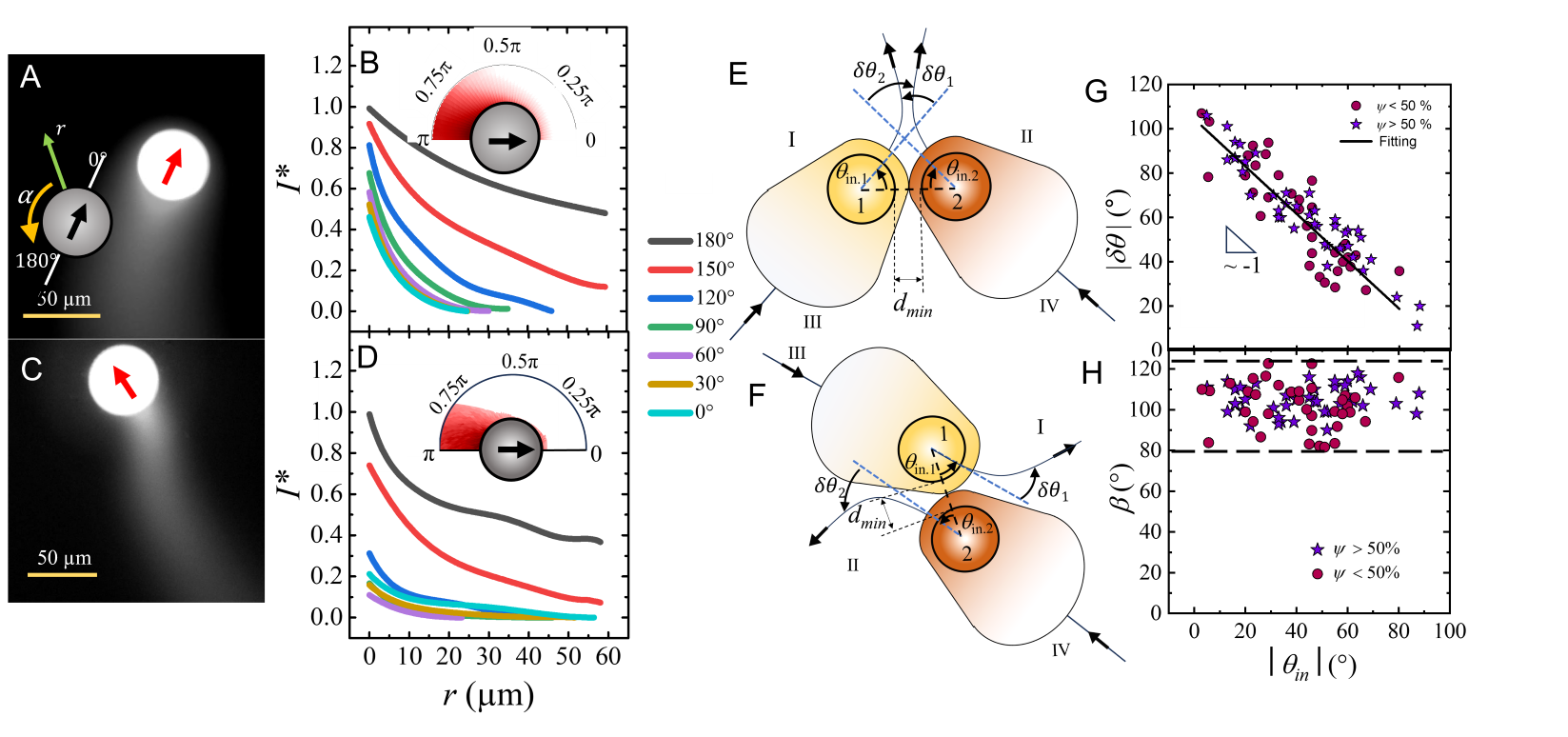}
\caption{ Fluorescence micrograph (greyscale) of an active 5CB droplet in aqueous 6 wt.$\%$ TTAB surfactant solution (A) with 1 wt.$\%$ PEO (Polyethylene oxide 8000 $kDa$ molecular weight) (C) with just water. The inset displays a schematic illustrating the measurement of fluorescent intensity ($I^{*}$) in the radial direction ($r$) at angular coordinates ($\alpha$) ranging from $0^\circ$ to $180^\circ$. Radial variation in fluorescence intensity $I^{*}$ around an active droplet starting from the droplet surface, at different angular positions (B) in 1 wt.$\%$ PEO and (D) in just water. Maximum intensity values ($I_{max}$) for 1 wt.\% PEO and water in recorded 8-bit images are 122.65 a.u. and 35.01 a.u., respectively, while minimum intensity values ($I_{min}$) are 0 a.u. for both cases. Here a.u. stands for arbitrary units. Schematic illustrating collision of the plume of filled micelles in droplet-droplet interaction for the case of (E) $\psi>50\%$ and (F) $\psi<50\%$. The yellow-colored droplet in (E) and (F) serves as the probe droplet, forming the basis of the discussion. Schematic also illustrates the approaching angle ($\theta_\text{{in.}}$), change in droplet orientation ($\delta\theta$), and minimum surface-to-surface distance between interacting droplets ($d_{\text{min.}}$) fo the two cases. (G) Variation in $|\delta\theta|$ with respect to $|\theta_\text{{in.}}|$. (H) Variation in $\beta$ with respect to $|\theta_\text{{in.}}|$.} 
\label{Figure3}
\end{figure*}

To gain better insights into how the self-generated chemical field affects the interaction between a pair of droplets, we analyze the chemical field distribution surrounding an isolated active droplet. This is accomplished by infusing the droplet with oil-soluble dye Nile Red. Since the droplet self-propulsion is accompanied by a continuous release of filled micelles through micellar solubilization, doping the droplet with dye enables visualization and estimation of the chemical field of filled micelles around the droplet \cite{hokmabad2022chemotactic,jin2017chemotaxis,dwivedi2023mode}. Fluorescence microscopy, with a laser at a wavelength of 560 nm, allowed us to both visualize and measure the fluorescent intensity $I(r,\alpha)$, extending radially from the droplet surface at various polar angles, $\alpha$, measured with respect to the direction of propulsion, as depicted in the inset of Fig. \ref{Figure3}A. Using the highest ($I_{max}$) and the lowest ($I_{min}$) values of intensity $I(r,\alpha)$, we computed $I^{*}= \frac{I(r,\alpha)-I_{min}}{I_{max}-I_{min}}$, which, as illustrated in Fig. \ref{Figure3}B, demonstrates a monotonic decay with radial distance ($r$) for an active droplet at different angular positions ($\alpha$). In the front region of the droplet (such as $\alpha = 0^{\circ}$, $30^\circ$, $60^\circ$, and $90^\circ$), the decay length ($\lambda(\alpha)$) of $I^{*}$, ranges from $\sim$ 25 $\mu$m to 35 $\mu$m. However, in the rear region (at angles beyond $90^\circ$, such as $120^\circ$, $150^\circ$, and $180^\circ$), $I^{*}$ does not decay completely, indicating an extended presence of filled micelles. The asymmetric distribution of filled micelles is expected due to the advection-driven transport of filled micelles towards the rear end of the droplet, forming an extended trail (see Fig. \ref{Figure3}A). It is to be noted that the concentration field in the anterior region for a droplet in PEO + TTAB aqueous solution is significantly higher compared to that for a droplet in PEO-free aqueous TTAB solution (see Fig. \ref{Figure3}C,D). Thus, an isolated self-propelling droplet has an intrinsic asymmetry in its chemical field, the strength of which is dictated by its $Pe$. This intrinsic asymmetry in the chemical field will be shown to play a crucial role in the pair-wise interactions. In the presence of PEO, droplets exhibit a lower $Pe$, wherein the relatively weaker advection of the surrounding fluid is unable to sweep away the filled micelles towards the rear of the droplet, with as much strength as droplets swimming at higher $Pe$ in PEO-free aqueous TTAB solution. Hence, in PEO solution, whenever two droplets approach in any orientation, the accumulation of filled micelles in the intermediate region significantly hampers the supply of fresh surfactants \cite{morozov2020adsorption}. Consequently, in line with the negative chemotactic behavior \cite{jin2017chemotaxis}, the droplets alter their paths in search of fresh fuel, thereby moving apart and reinstating the necessary interfacial polarity for their motion. 


The addition of high molecular weight polymer is known to impart viscoelastic properties to the host media \cite{bird1987dynamics}. However, as shown in our previous work \cite{dwivedi2023mode}, introducing 1 wt.\% PEO (8000 kDa) results in a low Deborah number ($De=0.04$), indicating that the time scale of droplet dynamics is longer than the relaxation time scale of the surrounding polymer. As a result, the PEO solution behaves as a Newtonian fluid with increased viscosity. Increasing the polymer and surfactant concentration raises $De$, at which point viscoelastic effects on the propelling droplets can cause deformation from their spherical shape \cite{dwivedi2023deforming}. Since no deformation was observed in the current study, the impact of viscoelasticity is deemed negligible. To further confirm that the repulsive scattering behavior in pairwise interactions is independent of viscoelastic effects from the polymer, we conducted additional experiments in a micelle-rich aqueous TTAB solution (see Materials and Methods). In this environment as well, droplets moved at a slower speed ($\sim$ 6 $\mu$m s$^{-1}$) compared to their speed in water ($\sim$ 30 $\mu$m s$^{-1}$), leading to a lower $Pe$. Similar to the behavior observed in the 1 wt.\% PEO solution, droplets in the micelle-rich environment exhibited puller swimming mode, drawing fluid from both front and rear ends with comparable repulsive pair interactions.

\begin{figure}
\includegraphics[scale=0.059]{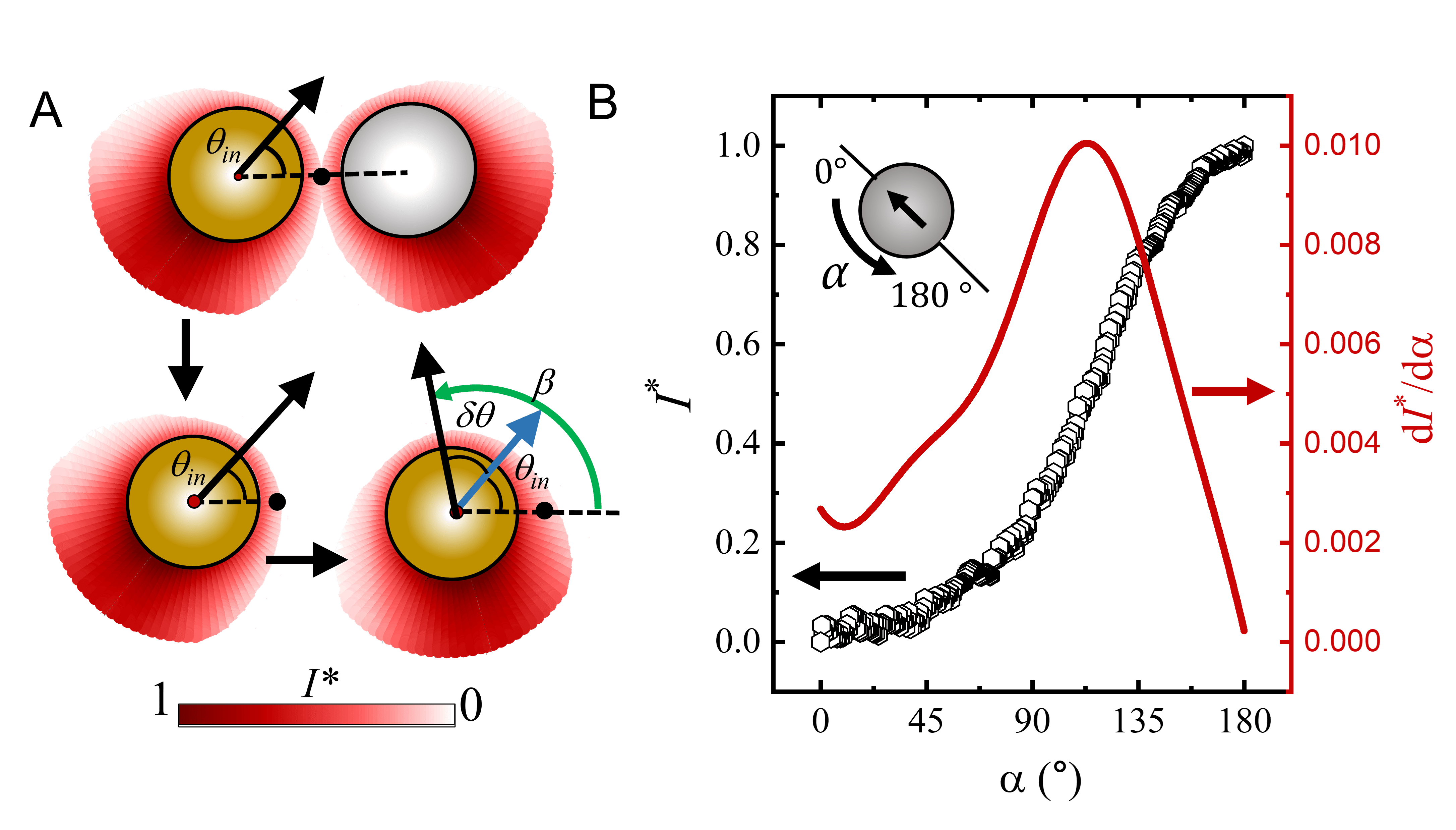}
\caption{\label{Fig:epsart}(A) Schematic depicting the impact point where filled micelles from interacting droplets meet. Further, the change in the droplet direction in the event of encounter of filled micelles from partner droplet at impact point is depicted in the bottom panels. Black and blue arrows represent the direction of droplet movement before and after deflection, respectively. Red colored plume around droplet represents the value of $I^{*}$ ranging from 0 to 1. (B) Variation in $I^{*}$ and $\frac{dI^{*}}{d\alpha}$ with respect to $\alpha$ along the droplet interface in the case of PEO+TTAB solution.}
\label{Figure4}
\end{figure}

\begin{figure*}
\centering
\includegraphics[scale=0.06]{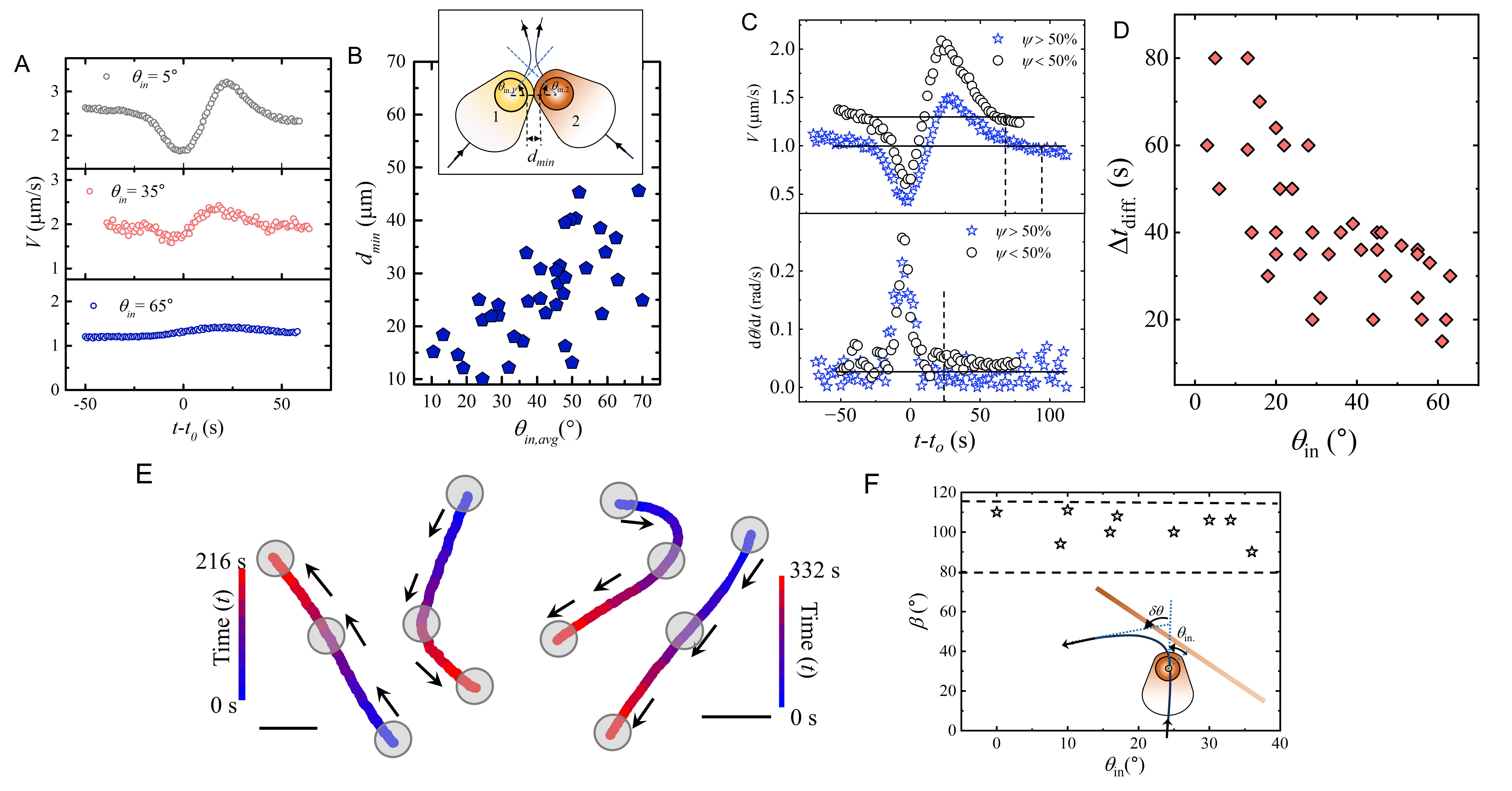}
\caption{(A) Evolution in droplet speed $V$ during pairwise interaction for different cases of $\theta_\text{{in.}}$. The $t_{0}$ is the reference time representing the moment when the distance between the surfaces of the droplets reaches $d_{\text{min.}}$. (B) Variation in $d_{\text{min.}}$ with respect to $\theta_\text{{in,avg}}$. Schematic in the inset illustrates $d_{\text{min.}}$. (C) Evolution in droplet speed $V$ and the rate of change in droplet orientation$\left(\left|\frac{d\theta}{dt}\right|\right)$ during pairwise interaction in case of $\psi>50\%$ and $\psi<50\%$. (D) Variation in $\Delta t_\text{{diff.}}$ with respect to the $\theta_\text{{in.}}$. (E) Representative trajectories of droplet interaction with trail/wake of other droplet in case of aqueous 1 wt.\% PEO solution. Scale bars indicate a length of 100 $\mu$m. (F) Variation in $\beta$ with respect to the $\theta_\text{{in,trail}}$ in droplet-trail interactions in case of aqueous 1 wt.\% PEO solution. Inset to the Figure illustrates the definition of angles.
}
\label{Figure5}
\end{figure*}

We now discuss the differences in the pair-wise interactions observed at higher and lower $\psi$ values. Illustrated in Fig. \ref{Figure3}E is a schematic representative of the closest approach (at $d_{\text{min.}}$) of a pair of droplets with $\psi$ $>$ 50 $\%$. Due to the accumulation of filled micelles, the droplets are prevented from direct contact; the overlap of their chemical plumes generates a chemo-repulsive interaction. Following their interaction, it is anticipated that the droplets will move towards regions not already occupied by filled micelles, thereby re-establishing a supply of fresh surfactants. If we identify the yellow droplet on the left as the probe droplet, it can then move away from filled micelle zones, to either region \emph{I} or \emph{II}. Given that both the droplets are identical and their approach is symmetric, their responses are also expected to be symmetric. Should the probe droplet head towards region \emph{II}, its counterpart is expected to proceed to region \emph{I}. This, however, would lead to overlapping paths, which is prohibited. Consequently, the probe droplet is compelled to move towards region \emph{I}, with its partner heading to region \emph{II}, thus creating trajectories that are reflective of one another. Similarly, Fig. \ref{Figure3}F displays a schematic representation of the droplets at $d_{\text{min.}}$ with a lesser potential overlap (i.e., $\psi<50\%$). Considering the filled micelles in regions \emph{III} and \emph{IV}, the probe droplet has the option to advance towards region \emph{I} or \emph{II}. Should the probe droplet head toward region \emph{II}, leading the partner droplet to choose region \emph{I}, their paths would intersect. Hence, the droplets are likely to move towards regions \emph{I} and \emph{II}, respectively, resulting in paths that do not mirror each other. As depicted in the schematic, we determine the incoming angle ($\theta_\text{{in.}}$), formed between the droplet velocity vector and the line joining the centers of the two droplets at $d_\text{{min}}$. We limit our focus on interactions with $\theta_\text{{in.}}$ $\le$ $90^\circ$ for both droplets at their minimum separation, $d_{\text{min.}}$. Further, we also determine $\delta\theta$, which is a measure of the change in orientation of a droplet's velocity vector due to the interaction. Fig. \ref{Figure3}G illustrates the variation in $|\delta\theta|$ with respect to $|\theta_\text{{in.}}|$, for all observed pair encounters. Notably, regardless of whether $\psi$ is above 50\% or below 50\%, $|\delta\theta|$ shows an inverse correlation with $|\theta_\text{{in.}}|$. This leads to the quantity, $\beta\equiv|\theta_{in}| + |\delta\theta|$ being nearly a constant, irrespective of changes in $|\theta_\text{{in.}}|$, as depicted in Fig. \ref{Figure3}H. Here, $\beta$ represents the angle between the velocity vector of droplet after deflection and the center-to-center line (Fig. \ref{Figure4}A), and its range falls within 90$^\circ$ to 120$^\circ$. 

We now return to the measurement of radial decay of $I^{*}$ for different angular position ($\alpha$) around an active droplet. Fig. \ref{Figure3}B demonstrates that the decay length of $I^{*}$ remains nearly the same at its front ($\alpha$ = 0$^\circ$ to $\alpha=$ 90$^\circ$). As a result, the accumulation of filled micelles at the front of the droplet assumes a nearly circular form. Hence, when the plumes of the droplets touch each other, with both $|\theta_\text{{in.}}|$ $<$ 90$^\circ$, the contact point is located along the line connecting their centers. Additionally, $|\theta_\text{{in.}}|$ and $\beta$ can be identified as the angles between the incoming and outgoing velocity vectors relative to the line joining the droplet center and its impact point, respectively (see Fig. \ref{Figure4}A). To understand the invariance of $\beta$ with respect to $|\theta_\text{{in.}}|$, we plot $I^{*}$ as well as $\frac{\text{d}I^{*}}{\text{d}\alpha}$ as a function of angular position $\alpha$ near the interface of an isolated moving droplet (see Fig. \ref{Figure4}B). The plot reveals a monotonic increase in $I^{*}$ from the front (0$^\circ$) to the rear of the droplet (180$^\circ$), consistent with higher concentration of filled micelles in the droplet's wake compared to its front. For $\alpha$ less than 90$^\circ$, $I^{*}$ increases gradually. However, between 90$^\circ$ and 120$^\circ$, $I^{*}$ rises sharply, indicating a rapid accumulation of filled micelles in this region. Beyond 120-130$^\circ$, rate of increase in $I^{*}$ again falls, and it eventually plateaus about 180$^\circ$. This distribution suggests the existence of a transition zone between 90$^\circ$ and 120$^\circ$, marking a sudden shift from lower to higher concentrations of filled micelles near the droplet's surface. This polar asymmetry in the filled micelle concentration ensures sustained propulsion of an isolated droplet in the same direction, until influenced by external concentration fluctuations. During a pair interaction, when the plumes of the filled micelles touch each other, the inherent angular asymmetry of the filled micelle concentration is disrupted at the impact point and a bilateral asymmetry about the propulsion direction is also introduced, as illustrated in schematic shown in Fig. \ref{Figure4}A. Although $Pe$ is low, the diffusion timescale needed for the micelles to diffuse and reinstate the droplets' intrinsic chemical polarity ($\tau_{\text{{diffuse}}}= \frac{R^2}{D}$ $\sim$ 100 s) is longer than the timescale of droplet motion ($\tau_{\text{translation}} = \frac{R}{V}$ $\sim$ 25 s). This forces the droplets to re-align their velocity vector by approximately $90^\circ$ to $120^\circ$ relative to the line joining the droplet centers at $d_{min}$, where the contact point is located. This adjustment eventually restores the original asymmetry of their surrounding chemical field and the steady swimming state, which is the stable state for the given $Pe$.

The magnitude of this deflection depends upon the severity of the penalty in chemical asymmetry enforced during the interaction. Therefore, lower values of $\theta_\text{{in.}}$ result in higher $\delta\theta$, and vice-versa. Hence, irrespective of $\theta_\text{{in.}}$, $\beta$ consistently lies within the range of 90$^\circ$ to 120$^\circ$. For the case of droplets interacting with high overlap ($\psi>50$$\%$) and mirrored trajectory, this manifests in droplets diverging at same angle independent of their approach orientations. In 2020, Lippera \emph{et al.} forecasted that regardless of the initial approach angle, droplets alter their swimming direction such that their departing velocity vectors form an angle that is nearly constant $\sim$ 60$^\circ$ (equivalent to $\beta$ $\sim$ 120$^\circ$) \cite{lippera2021alignment}. Our experimental observations showcase a striking correspondence with the predictions made by Lippera \emph{et al.}, thereby reinforcing the validity of their model. Further, our study demonstrates that this observation remains true, even for the case of non-symmetric collisions (i.e., for $\psi<50$$\%$). 

Fig. \ref{Figure5}A demonstrates the evolution of droplet speed, $V$, during pair-wise interaction for different values of $\theta_\text{{in.}}$. It is evident that a droplet with lower values of $\theta_\text{{in.}}$ experiences a greater relative magnitude of change in speed, leading to a higher overall change in momentum. This momentum change stems from the chemo-repulsive interaction due to the build-up of filled micelles between the surfaces of the droplets. With a reduction in the gap between the droplet surfaces, the micellar accumulation builds up, resulting in a higher repulsion. Consequently, for lower values of $\theta_\text{{in.}}$, wherein the droplets have to undergo maximum deflection, they approach closer (i.e., attain a lower $d_{\text{min.}}$), and experience stronger repulsion. This is supported by an increasing trend in $d_{\text{min.}}$ with respect to $\theta_\text{{in,avg}}=\frac{\theta_\text{{in,1}}+\theta_\text{{in,2}}}{2}$, where $\theta_\text{{in,1}}$ and $\theta_\text{{in,2}}$ are the incident angles of the two interacting droplets, shown in Fig. \ref{Figure5}B.  Fig. \ref{Figure5}C reveals that the duration ($\Delta t$) over which the translational velocity $V$ deviates from its initial pre-collision value, exceeds the $\Delta t$ during which the rate of change in angular speed $\left(\left|\frac{d\theta}{dt}\right|\right)$ is nonzero. This difference, denoted as $\Delta t_\text{{diff.}}$, suggests that even after the cessation of droplet rotation, the sustained accumulation of the chemical field maintains an elevated translational speed of the rebounding droplets. This argument is supported by the decreasing behavior of $\Delta t_\text{{diff.}}$ with respect to $\theta_\text{{in.}}$, as shown in Fig. \ref{Figure5}D. For lower $\theta_\text{{in.}}$, due to a higher velocity component along the droplet center-to-center line, the higher extent of overlap of the chemical fields leads to a prolonged effect on the translation motion, resulting in a higher $\Delta t_\text{{diff.}}$. On the other hand, since the extent of overlap of plumes is lesser in cases with higher $\theta_\text{{in.}}$, the effect wanes off with smaller $\Delta t_\text{{diff.}}$.

Next, we discuss the interaction of a droplet with the chemical trail left by another droplet for which $\theta_\text{{in.}}$ $>$ $90^\circ$. Fig. \ref{Figure5}E depict representative trajectories for such interactions. Fig. \ref{Figure5}F presents the variation of $\beta$ (= $\theta_\text{{in.}}$ + $\delta \theta$), as described in the schematic shown in the inset, with varying incident angle, $\theta_\text{{in.}}$. Similar to our observations for droplet-droplet interactions, regardless of their initial approach direction, droplets consistently rebound with $\beta$ values ranging between 90$^\circ$ and 120$^\circ$. This observation further underscores the conclusion that the rebounding behavior of the droplet is mainly decided by the intrinsic chemical polarity of the self-propelling droplets. 

\subsection{\label{sec:level2}Interactions at high $Pe$}

\begin{figure*}[tbhp]
\centering
\includegraphics[scale=0.1]{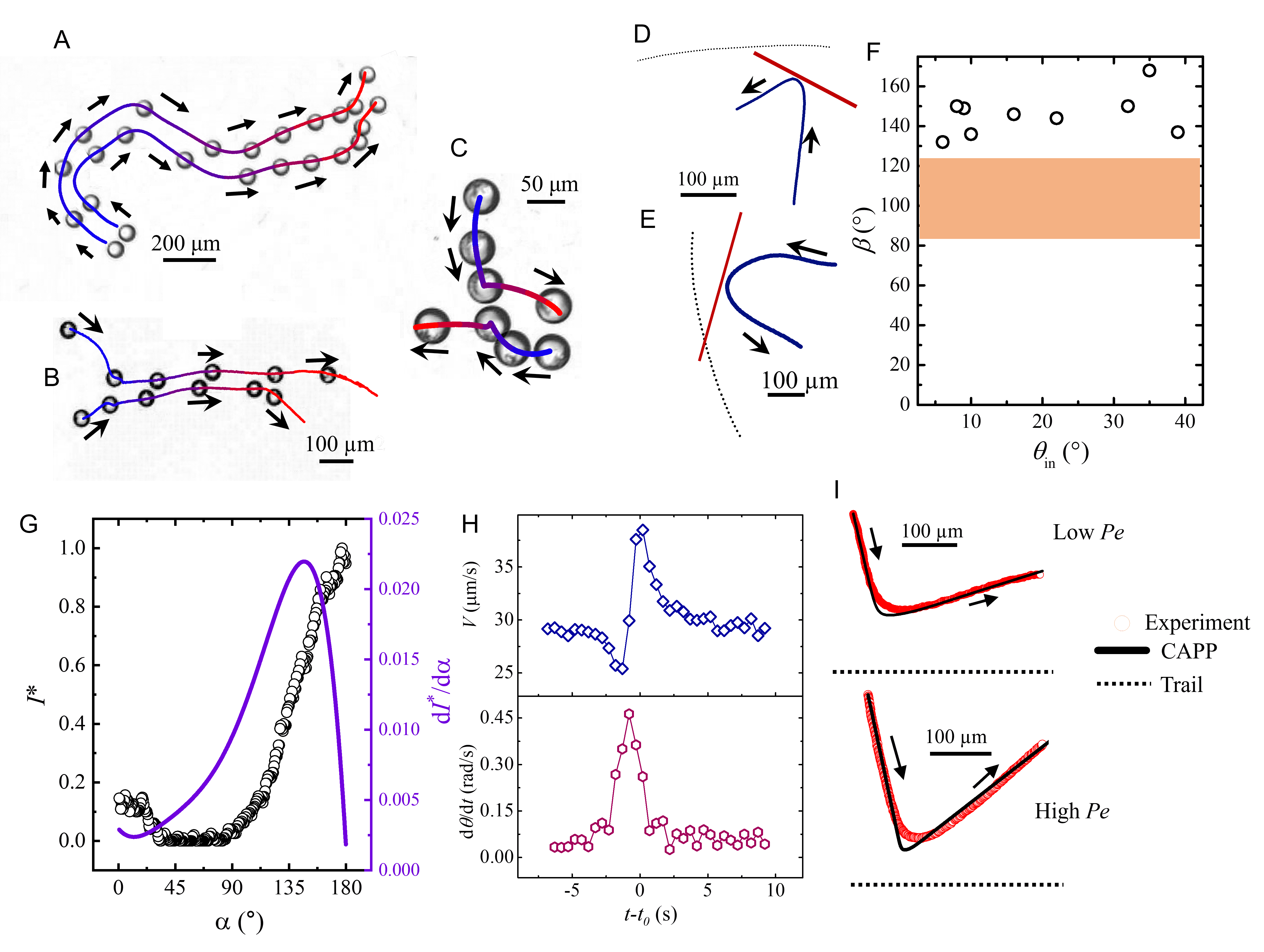}
\caption{ Active 5CB droplets swimming in aqueous 6 wt.$\%$ TTAB surfactant solution (A-C) Representative trajectories of pairwise interactions. (D,E) Representative trajectories of droplets interacting with chemical trails left behind by other swimming droplets. Dotted black line represents the center line of the chemical trail and solid red line is the anticipated reflective line. (F) Variation of $\beta$ with incident angle $\theta_\text{{in}}$ in case of water. Definition of angles bear resemblance to those illustrated in Fig.\ref{Figure5}F. (G) Variation in $I^{*}$ and $\frac{dI^{*}}{d\alpha}$ with respect to the $\alpha$ along the droplet interface in case of water.(H) Representative variation of $V$ and $\left|\frac{d\theta}{dt}\right|$ with time in case of water. $t_{0}$ represents the reference time at which the distance between the droplet and the center line along the trail is minimum. (I) Representative trajectories showing the droplets rebounding during the interaction with the trail/wake of other droplet. Black dotted lines represents the center line of the chemical trail left behind by other swimming droplets. Overlaid black line on droplet trajectories represents the trajectories obtained using CAPP model. }
\label{Figure6}
\end{figure*}

The observations in the previous section emphasize the significance of chemical field-driven repulsive interactions between droplet-droplet/wake at low Péclet numbers ($Pe$). To contrast these findings with pair-wise interactions at relatively higher $Pe$ values, where the hydrodynamic effects are expected to dictate the droplet interactions, we explore interactions of 5CB droplets in PEO-free aqueous TTAB solution, wherein $Pe$ $\sim$ 75. We observe $\sim$ 20 pair interactions, and a few representative trajectories are shown in Fig. \ref{Figure6}A-C. 
In contrast to the behavior observed at lower $Pe$ values, it is observed that droplets approach very close, occasionally even making sustained contact. On approaching from the same direction, they typically move in unison for sometime before separating. This synchronous motion can occur with multiple formations, such as (a) side-by-side (Fig. \ref{Figure6}A), where droplets stay close to each other's equatorial position - a conFiguration typical when $\theta_\text{{in.}}$ is less than or equal to 90$^{\circ}$; and (b) staggered (Fig. \ref{Figure6}B), where one droplet trails the other at an angle, commonly observed when a droplet's $\theta_\text{{in.}}$ is between 90 and 180$^{\circ}$. These conFigurations often shift from side-by-side to staggered with $d_{min}\sim$10 $\mu$m before the droplets part ways. Further, when droplets meet head-on with low $\theta_\text{{in.}}$, they almost make contact with $d_{min}=$ 0, pause briefly, and then move apart, as illustrated in Fig. \ref{Figure6}C. Since pairwise interactions do not exhibit scattering behaviour at higher $Pe$, the value of $\beta$ has not been determined in these cases.

Unlike low $Pe$ systems, the absence of strong repulsion indicates relatively inferior chemical interactions. Better understanding of these observations is extracted by investigating the angular variation of the chemical field around an isolated droplet navigating through an aqueous TTAB solution, as reflected by the normalized fluorescence intensity $I^{*}$ (refer to Fig. \ref{Figure3}C,D). In the anterior region of the droplet (i.e., at angles of $0^\circ$, $30^\circ$, $60^\circ$, and $90^\circ$), $I^{*}$ is remarkably lower compared to the case of droplet at low $Pe$. However, in the rear region (beyond $120^\circ$), the intensity values increase sharply, implying that the presence of filled micelles is confined to a narrow wake region of the active droplet, while being almost absent at its front and sides. Notably, the intensity levels at angles of $150^\circ$ and $180^\circ$ are actually lower at higher $Pe$ values, than those observed in lower $Pe$ settings. This difference is attributed to the stronger advection at higher $Pe$, which effectively convects away the filled micelles to the droplet's rear, thereby diminishing their overall concentration. This reduction in the concentration of filled micelles leads to a decreased chemotactic repulsion among the droplets. As such, the observation of attraction, albeit weak, is an evidence of significant influence of hydrodynamics in the pair-interactions as seen in Fig. \ref{Figure6}A-C. At $Pe$ $\sim$ 75, these droplets swim with weak pushing gait, as confirmed through PIV experiments (see Fig. \ref{Figure1}C). In 2010, Götze \emph{et al.}\ theoretically demonstrated that a pair of strong pusher squirmers to indefinitely maintain a side-by-side conFiguration \cite{gotze2010mesoscale}. However, introducing orientational fluctuations was found to induce separation. Our findings confirm the predictions made by Götze \emph{et al.}, highlighting the significance of droplet hydrodynamics in their pairwise interactions at higher $Pe$. Further, the local time-dependent fluctuations in surfactant concentration at the interface introduce the orientation fluctuations in droplet motion \cite{suda2021straight} which result in their eventual separation. During their head-on approach, the absence of strong chemotactic repulsion facilitates very low $d_{min}$. This is despite the outwardly emanating streamlines from the front of the droplets, which theoretically ought to hydrodynamically repel the droplets. However, it is only once they approach very close that the short range ($\sim$ $\frac{1}{r^{3}}$) hydrodynamic interaction results in their separation. Overall, our observations are in line with prior theoretical investigations that suggest pusher squirmers to move together due to hydrodynamic attraction \cite{gotze2010mesoscale,guan2022swimming,ouyang2019hydrodynamic,kanevsky2010modeling}. They also align with recent experimental evidence from studies on micellar solubilization-based active droplets in surfactant solutions, where hydrodynamic interactions have been observed to result in the formation of metastable lines \cite{thutupalli2018flow} and clustering \cite{hokmabad2022spontaneously}.

Thus far, we have focused on droplet-droplet interactions at higher $Pe$; we next shift our attention to droplet-wake interactions under similar conditions. Fig. \ref{Figure6}D,E depict representative trajectories of droplet-wake interactions at high $Pe$, with dotted lines indicating the wake of a droplet, and solid blue lines representing the interacting droplet trajectories. Fig. \ref{Figure6}F presents the variation of $\beta$ (= $\theta_\text{{in.}}$ + $\delta \theta$), with varying incident angle $\theta_\text{{in.}}$ for a few such interactions. Similar to our observations at low Péclet numbers ($Pe$), we observe that regardless of their initial approach direction, droplets consistently rebound with $\beta$ values ranging between 130° and 150°. It is noteworthy to emphasize that the range of $\beta$ values for aqueous TTAB solution (shown by markers) exceeds that of TTAB+PEO aqueous solution (shaded region in Fig. \ref{Figure6}F). This observation aligns with the chemical rebounding phenomenon, which is influenced by the inherent chemical polarity of the droplets. This difference in inherent chemical polarity between 5CB droplets swimming with low and high $Pe$ values is indicated by the angular variation of $I^{*}$ shown in Fig. \ref{Figure4}B and Fig. \ref{Figure6}G. For aqueous TTAB solution, Hokamabad \emph{et al.} \cite{hokmabad2022chemotactic}, reported that active droplets ``reflect" when interacting with the chemical wake of other droplets. Numerical analysis using the chemically active polar particle (CAPP) approach was further illustrated to match with the experimental findings. While these computational analyses agree with seemingly reflective trajectories, they do not account for the droplet's inherent $Pe$-dependent chemical polarity. Fig. \ref{Figure6}D-E highlights that the expected reflection planes (solid red lines) do not coincide with the droplet wake ( dotted black lines). This discrepancy indicates that the droplets' behavior can be exceedingly more complex than mere reflection about the wake. Unlike low $Pe$ case, for these interactions, both angular speed $\left(\left|\frac{d\theta}{dt}\right|\right)$ and translation speed, $V$, return to their pre-collision value at around same time, i.e., $\Delta t_\text{{diff.}}$ $\sim$ 0 (see Fig. \ref{Figure6}H). This is again attributed to the relatively weaker strength of chemical interactions at higher $Pe$. 

The balance between solute advection and diffusion can be used to determine the critical non-dimensional distance at which a self-propelling droplet detects the presence of the hydrodynamic or chemical field of another droplet. This critical distance is given by $r_c/R \sim Pe^{-1}$, where $r_c$ is the critical radial distance from the droplet center, $R$ is the droplet radius. The time scale for a droplet to traverse this critical distance, $r_c$, can be estimated as $r_c/V \sim D/V^2$. Experimental time scales observed from \ref{Figure5}C ($\sim125$ s for low $Pe$) and \ref{Figure6}H ($\sim7$ s for high $Pe$) appear to align with these scaling estimates. However, varying the $Pe$ across several orders of magnitude alters the droplet's swimming mode, consequently changing its hydrodynamic signature. Therefore, a more comprehensive analysis is required to fully understand how the time scale depends on $Pe$.

Finally, we use the CAPP model \cite{saha2014clusters} to predict the droplet-wake interaction for the case of both low and high $Pe$. The wake is assumed to be a quasi-static 1D diffusing concentration field in the $y$-direction, with the source located at $y=0$. The evolution of the droplet position and its orientation as it interacts with the wake is obtained from the Langevin dynamics (ignoring random fluctuations) as 
\begin{subequations}
    \begin{equation}
    \dot{x}=V\cos\theta,
    \end{equation}
    \begin{equation}
    \dot{y}=V\sin\theta +\frac{\alpha c_0y}{\sqrt{2\pi}\left(2D\Delta t\right)^{3/2}}\exp\left(-\frac{y^2}{4D\Delta t}\right),
\end{equation}
\begin{equation}
    \dot{\theta}=\frac{\Omega c_0y}{\sqrt{2\pi}\left(2D\Delta t\right)^{3/2}}\exp\left(-\frac{y^2}{4D\Delta t}\right)\cos\theta.
\end{equation}
\end{subequations}

\begin{table}
\caption{\label{tab:table1}\textbf Comparison of dimensionless coupling constants obtained for fitting the X-Y trajectories using the CAPP model.  }
\begin{ruledtabular}
\begin{tabular}{lcr}
$Pe$&\text{$\frac{\Omega c_{0}}{VR}$}&$\frac{\alpha c_{0}}{VR^2}$\\
\hline
Low (TTAB+PEO solution) & 12 & 12.8\\
High (TTAB solution) & 60 & 8\\
\end{tabular}
\end{ruledtabular}
\end{table}


The model has two fitting parameters, $\Omega c_{0}$ and $\alpha c_{0}$, which represent the strength of effective torque and the effective translational deceleration experienced by the droplet, respectively. To ensure a valid comparison between the low and high $Pe$ cases, we normalize the coupling constants with respect to $VR$ and $VR^{2}$, respectively. The normalized coupling constants are listed in Table I. The values indicate that at higher $Pe$, droplets experience a higher torque $\sim$ 5 times than at lower $Pe$, which is consistent with the higher $\beta$ values observed for droplets at higher $Pe$ in order to re-establish their inherent chemical polarity. Further, droplets at lower $Pe$ experience a more pronounced translational deceleration/acceleration, consistent with increased accumulation of filled micelles in intermediate region, leading to chemo-repulsive interactions. The predicted and experimental droplet trajectories for high and low $Pe$ droplet-wake interactions are depicted in Fig. \ref{Figure6}I, and are found to be in good agreement. 

\section{\label{sec:level1}CONCLUSIONS}
This study presents a comprehensive investigation into the pair-wise interactions of chemically active droplets swimming at different Péclet numbers ($Pe$). At low $Pe$ values, the observed pair-wise interactions between approaching droplets showcase a distinctive scattering behaviour, wherein, after reaching a minimum distance the droplets rebound without engaging in direct physical contact. This rebounding/scattering phenomenon is attributed to the dominant influence of the accumulation of chemical field emanating from the filled micelles. Due to weaker advection, the influence of the hydrodynamic field is minimal at low $Pe$, thereby allowing chemo-repulsive interactions to outweigh the contribution from their hydrodynamic interactions. The findings of this study provide excellent experimental evidence supporting the conclusions drawn from the theoretical studies by Michelin and co-workers \cite{lippera2020collisions,lippera2021alignment}. In contrast to low $Pe$ conditions, droplets swimming at higher $Pe$ demonstrate dominance of hydrodynamic fields in their pair-wise interactions, thus allowing droplets to attract and even sustain contact. Our study also demonstrates that, independent of the underlying $Pe$, in both cases, droplet-wake interactions are influenced primarily by the repulsive chemical interactions. The chemo repulsive interaction is always dictated by the inherent chemical polarity of the deflecting droplet, as determined by its $Pe$. This study provides compelling evidence highlighting the significant influence of both hydrodynamic and chemical fields on the dynamics of active droplets during their pair-wise interactions across a wide range of Péclet numbers ($Pe$). The observed results offer valuable insights that contribute to a deeper understanding of multibody interactions for low $Pe$, and high $Pe$ systems within active droplet systems. Moreover, the implications of this research extend to the potential for controlling and optimizing such interactions under diverse $Pe$ conditions.  


\section*{ACKNOWLEDGEMENTs}
Authors acknowledge the funding received from the Science and Engineering Research Board (ECR/2018/000401 and CRG/2022/003763), and from Department of Science and Technology, India (SR/FST/ETII-055/2013)

\section*{AUTHORS DECLARATION}
\subsection*{Conflict of Interest}
The authors have no conflicts to disclose.

\bibliography{aipsamp}

\end{document}